# CHARON'S RADIUS AND DENSITY FROM THE COMBINED DATA SETS OF THE 2005 JULY 11 OCCULTATION

by

M. J. Person[1], J. L. Elliot[1,2,3], A. A. S. Gulbis[1], J. M. Pasachoff[4], B. A. Babcock[5],

S. P. Souza[4], and J. Gangestad[4]

[1]*Department of Earth, Atmospheric, and Planetary Sciences, Massachusetts Institute of Technology, Cambridge, MA 02139-4307, USA*

[2]*Department of Physics, Massachusetts Institute of Technology, Cambridge, MA 02139-4307, USA*

[3]*Lowell Observatory, Flagstaff, AZ 86001, USA*

[4]*Hopkins Observatory, Williams College, Williamstown, MA 01267-2565, USA*

[5]*Physics Department, Williams College, Williamstown, MA 01267-2565, USA*

E-mail queries: contact mjperson@mit.edu

Short Title: "Charon's Radius and Density"

ABSTRACT

The 2005 July 11 C313.2 stellar occultation by Charon was observed by three separate research groups, including our own, at observatories throughout South America. Here, the published timings from the three data sets have been combined to more accurately determine the mean radius of Charon: 606.0 ± 1.5 km. Our analysis indicates that a slight oblateness in the body (0.006 ± 0.003) best matches the data, with a confidence level of 86%. The oblateness has a pole position angle of 71.4º ± 10.4º and is consistent with Charon's pole position angle of 67º. Charon's mean radius corresponds to a bulk density of 1.63 ± 0.07 g/cm$^3$, which is significantly less than Pluto's (1.92 ± 0.12 g/cm$^3$). This density differential favors an impact formation scenario for the system in which at least one of the impactors was differentiated. Finally, unexplained differences between chord timings measured at Cerro Pachón and the rest of the data set could be indicative of a depression as deep as 7 km on Charon's limb.

## 1. INTRODUCTION

Recent discoveries of large, presumably pristine objects in the Kuiper Belt, such as 2003UB$_{313}$, 2005FY$_{9}$, and 2003EL$_{61}$ (Brown et al. 2005a), provide sources of valuable data for modeling our solar system's formation and evolution. Unfortunately, the great distances to these objects (currently at 97 AU, 52 AU, and 52 AU, respectively) require large amounts of time on the largest telescopes for data collection. Measurements of their actual sizes are difficult, and estimates of their masses and densities (and therefore rock/ice mass ratios), key information for solar-system evolution models, are possible only for the binary systems.

As our understanding of the outer solar system has expanded, it has become clear that planet Pluto and its moon Charon, as well as Neptune's captured satellite Triton, are likely members of the same overall parent population as these more distant Kuiper Belt objects. Thus, the study of Pluto, Charon, and Triton can provide valuable insights into the greater population. Given the much smaller distance at the current location of its orbit (~31 AU in 2005) study of the Pluto-Charon system can provide greater detail than observations of its more distant brethren currently allow.

One fundamental uncertainty regarding the Pluto-Charon system in recent years has been the size of Charon itself, which contributes (cubed) to the corresponding uncertainty in its density. A lower limit on Charon's radius of 600 km was derived from the single chord observed during the 1980 stellar occultation (Walker 1980), which was later revised to 601.5 km (Elliot & Young 1991). A mean radius for Charon was then derived from the Pluto-Charon mutual occultation events in the 1980s with a value ranging from 591 ± 5 km to 628 ± 21 km, depending upon limb-darkening assumptions and data selections (Tholen & Buie 1990; Reinsch et al. 1994; Young & Binzel 1994). An opportunity to improve on these values occurred in 2005 when Charon occulted a 14$^{th}$ magnitude star designated C313.2 (UCAC2 26257135, McDonald & Elliot 2000).

The C313.2 event was observed from telescopes throughout South America by three different research groups: Southwest Research Institute (SwRI) – Wellesley College, a group led by the Paris Observatory, and our Massachusetts Institute of Technology (MIT)

– Williams College team. In Young et al. (2005), the SwRI-Wellesley team presented their initial results. As they had essentially one chord, only a lower limit on the radius could be presented. This lower limit of 589.5 ± 2 km was consistent with earlier results but did not improve upon previous work. In the 2006 January 5 issue of *Nature*, the other two groups presented their results of the observations. The MIT-Williams consortium, as reported by Gulbis et al. (2006), analyzed four chords obtained at three different locations, yielding a radius of 606 ± 8 km (with the relatively large error bar accounting for possible elliptical solutions). The group led by the Paris Observatory, as reported by Sicardy et al. (2006), analyzed three of their occultation chords based on a circular model for the limb to deduce a radius of 603.6 ± 1.4 km.

As the Gulbis et al. (2006) data set included one occultation chord that was in itself larger than the Sicardy et al. (2006) combined solution (605.12 ± 0.05 km versus 603.6 ± 1.4 km, respectively), there is an inconsistency between the two results. The difference in mean radii is not surprising, given the low number of degrees of freedom in any separate three- or four-chord fit. Particularly in cases where none of the chords is particularly central (such as solutions without the Las Campanas chords), no strong constraint on the shadow diameter is provided. It is therefore profitable to combine all three data sets into one analysis, the results of which are presented here. Independent analyses performed at both MIT and Williams College are consistent with these results (Gangestad 2006; Person 2006).

## 2. DATA

Of the parameters reported in the literature, three are required for each chord used in the analysis: (a) the immersion/emersion light times, (b) the formal errors on these light times, and (c) the geodetic position (longitude and latitude) of the telescope from which the measurements were made. The published values for these quantities for the C313.2 Charon occultation from all reported efforts are presented in Table 1.

INSERT TABLE 1

The accuracy of the immersion/emersion times is affected by two factors, the photometric calibration of the occultation light curve from full stellar signal to zero stellar signal and the choice of light level at which the occultation was measured. Gulbis et al. (2006) reported "geometric-limb times" (measured at the 25% light level for a monochromatic Fresnel-diffraction pattern) which corresponds to the location of a straight edge occulting a point source for a monochromatic Fresnel-diffraction model, averaged over the integration time of the data. This averaging effect proved to be much larger than the averaging resulting from the finite stellar diameter and finite wavelength passband of the detected light. Sicardy et al. (2006) reported occultation times using a fit for a straight edge convolved with diffraction, including the averaging from the finite diameter of the occultation star. Given the small size of the occultation star in the occultation plane, 0.63 km [an apparent diameter of 28 microacrsec (corresponding to 0.02 s of shadow movement) was measured by Gerard van Belle of the Palomar Test Bed Interferometer at our request], the Gulbis et al. (2006) and the Sicardy et al. (2006) definitions should be consistent. Young et al. (2005) reported simply "immersion and emersion times," which can be assumed by common usage to be somewhere between geometric-limb times and half-light times. This difference should be less than 0.05 seconds (one quarter of the data cycle time recorded at SOAR), which, when multiplied by the typical topocentric shadow velocities of 21.3 km/s, could result in ambiguities of approximately 1 km.

The formal errors on the occultation times reported by the several groups were low (Table 1), ranging from 0.001 s for immersion in the Las Campanas light curve at the Clay telescope (Gulbis et al. 2006) to 0.30 s reported for emersion of the San Pedro de Atacama light curve (Sicardy et al. 2006), with differences resulting from different noise levels in the data. Hence these timing errors should be suitable for weighting the data.

Factors such as seeing conditions and photon noise lead to random errors in the timings, which propagate via the fitting procedures into formal errors on Charon's radius. However, errors in the absolute timing calibrations introduce systematic errors into the results. The PHOT camera used by Young et al. (2005, personal communication), and the POETS cameras used by our MIT-Williams team (Gulbis et al. 2006; Souza et al. 2006) are both directly triggered by GPS timing signals. GPS timing signals, once properly locked by the ground receivers, can be as accurate as 10 ns – far improved over the several millisecond accuracy required for this analysis (Souza et al. 2006). The Gemini South chord reported by Gulbis et al. (2006) derives its time from Network Time Protocols (NTP), which should be accurate to milliseconds, although high or variable network latency can degrade this accuracy (Deeths & Brunette 2001). Sicardy et al. (2006) did not report on the accuracy of their timing sources beyond the calculated formal errors. Errors in absolute timing calibration enter directly (multiplied by 21.3 km/s for this event) as systematic errors in the immersion and emersion locations along the station chords.

The geodetic locations from Gulbis et al. (2006) were obtained by GPS surveys for all stations excluding Gemini South, the location of which was taken from the Astronomical Almanac (2003). Observing station locations from Sicardy et al. (2006) are used as reported. Since the location of the SOAR telescope was not reported in Young et al. (2005), it was taken from the Astronomical Almanac (2003). Errors in the geodetic locations of the stations translate directly into errors on the occultation shadow plane, but modern surveying techniques make errors larger than a kilometer quite unlikely.

## 3. FITTING PROCEDURE

The station geodetic coordinates and event times were used to plot the station locations in the *(f,g)* occultation shadow plane according to the methodology of Elliot et al. (1992; 1996). This plane is centered on the coordinates of the occultation star [J2000.0 RA: $17^h$ $28^m$ $55.^s0174$, Dec: –15° 00' 54.750'' (Zacharias et al. 2004)], with the *f*-axis pointing east and the *g*-axis pointing north. The ephemeris position of Charon [JPL Ephemeris DE413/PLU013 (Chamberlin 2005)] at the time of the occultation is also converted into *(f,g)* coordinates and then subtracted from corresponding station coordinates. This conversion provides occultation immersion and emersion locations, for each station, in a Charon-centered coordinate system. For the combined C313.2 data set, these occultation locations are displayed in Figure 1.

INSERT FIGURE 1

Once in this system, a model can be fit to the data points (which should all lie along the limb of Charon) using standard least-squares methods (Bevington & Robinson 1992). The squared residuals between the points and a Charon limb model must be minimized in the direction of the occultation chords, as it is this direction that the formal timing errors describe (Elliot et al. 2000a). Errors in the perpendicular direction can be caused by errors in the geodetic positions of the stations and should be treated separately. Photometric mis-calibrations of the reported light curves can result in errors along the chord direction.

Both circular and elliptical models were fit to the (*f,g*)-plane immersion and emersion data points, using as weights the inverse of the product of the shadow velocity and the formal timing errors, squared, to determine the radius and limb shape of Charon. These models assume overall limb roughness is less than the random timing errors in the data. The results of these fits are listed in Table 2.

INSERT TABLE 2

## 4. FIT RESULTS

Using all the data, the first fit (Fit 1 in Table 2) of a circular figure gives a radius of 606.03 ± 0.20 km. However, this fit has an unacceptably high reduced chi-square value. Examination of the residuals to this fit shows that the greatest contribution to the chi-square value is from the residuals of the SOAR and Gemini stations.

As seen in Table 1, three sets of stations yielded essentially coincident occultation chords: Cerro Armazones and Paranal, Las Campanas Clay and du Pont, and SOAR and Gemini. While the first two pairs of stations have consistent timing (for example, the independently measured immersion times at the Las Campanas stations differ by 0.018 ± 0.005 s with an expected difference of 0.018 s), the SOAR and Gemini chords are offset in time by approximately 0.7 sec. This timing difference results in a required combined residual of at least 15 km, regardless of the model used. Given the formal errors in the SOAR and Gemini chords, residuals of approximately 3 km should be expected. Attempts to minimize the combined residual places the model between the data points, increasing the residuals at the data points from the other stations and resulting in the poor quality of the fit.

Since there is no clear resolution to the timing discrepancy between the two stations, the Gemini South and SOAR chords are dropped entirely in Fit 2. An immediate improvement is seen in the reduced chi square (1.27), indicating that the other stations (when Gemini South and SOAR are excluded) have residuals consistent with their formal timing errors. Fit 2 is displayed as a solid circle overlying the data points in Figure 1. For this fit, the SOAR data points have residuals of –7.1 km and 0.1 km, while the Gemini South points have residuals of 5.9 km and –13.8 km. To illustrate these residuals better, an expanded view of Fit 2 and the Gemini South and SOAR data points is presented in Figure 2. Note that the sums of their residuals, –7.0 km and –7.9 km, respectively, are consistent to within 0.9 km. This consistency is expected, given that the two chords have nearly identical durations even though they are shifted with respect to each other in time. The consistency of the chord length measured at these two stations mutually corroborates the length of the chord at Cerro Pachón.

INSERT FIGURE 2

One interpretation of this distortion is as a systematic error in the timing of one of the two stations. In this case, it would seem likely that the SOAR data are more accurate than the Gemini, since the SOAR residuals are lower with respect to Fit 2. Fit 3 explores this by including all stations except Gemini South. Even in this case the SOAR residuals of –5.8 km and 1.2 km are still much larger that the ~0.2 km residuals expected from the reported 0.01 s formal errors on these points. This problem can be seen in the fitted model being distorted away from the other stations, resulting once more in an unacceptably high reduced chi square for Fit 3. These results imply that either (a) SOAR and Gemini South suffered from independent systematic timing errors, or (b) Charon's limb profile does not follow a simple circular model at that location.

To measure an overall deviation from a circular limb profile, the data are next fit with an elliptical model. The elliptical model fit, with all data included, is Fit 4 in Table 2. Again, the reduced chi-square value is unacceptably high, although significantly lower than some values obtained with the circular model. The oblateness of 0.027 is significant, although it is only slightly greater than twice its fitted error bar. The ellipse's pole position angle of 95.5º ± 7.8º does not correspond with the position angle of Charon's rotation axis (approximately 67º).

For the same reasoning as the circular fits, the data are next fit with an elliptical model excluding the Gemini and SOAR chords. This result is given in Fit 5, which is displayed as a dashed curve in Figures 1 and 2. The oblateness of this fit is relatively small, at 0.006, but again is twice its formal error bar. In this case, the position angle of the ellipse does correspond to Charon's rotation axis, though with an error bar of over twice the measured difference between the rotation axis and that of the ellipse. This fit results in an acceptable reduced chi-square value of 0.89, again indicating that the remaining stations have residuals consistent with their formal errors. Given the decrease in degrees of freedom from Fit 2 (11 in Fit 2; 7 in Fit 5), the reduced chi square of Fit 5 (1.27 in Fit 2; 0.89 in Fit 5) indicates that the elliptical solution could be a better fit than the circular model. That assessment is likely, but not certain, since the probability of a circular data set with Gaussian errors being seemingly more accurately described by an elliptical model is 14% for these degrees of freedom.

Given the reduced chi-square values of the fits, we find that Fits 2 and 5, in which the Gemini South and SOAR chords were excluded from the fitted solutions, are the best representations of Charon's global figure. These two solutions are in boldface in Table 2. Assuming a simple circular shape for Charon's limb profile, we find that Charon's mean radius is 606.01 ± 0.02 km (Fit 2). However, there are indications from both the results of Fit 5 and the irregularities of the Gemini and SOAR chords that the simple circular model is insufficient. Therefore, in the manner of Gulbis et al. (2006), we increase the error bar to include the mean radius of the best-fitting elliptical solution (geometric mean of the semimajor and semiminor axes). This results in an overall mean radius for Charon of 606.0 ± 1.5 km. We believe this result to be a significant improvement over previously reported measurements, as shown in Table 3.

INSERT TABLE 3

By combining this radius result with the most recent measurement of Charon's mass, 1.520 ± 0.064 x $10^{21}$kg (Buie et al. 2006), we find a bulk density for Charon of 1.63 ± 0.07 g/cm$^3$. Assuming densities for ice of 1.0 g/cm$^3$ and rock of 3.0 g/cm$^3$, the rock mass fraction of Charon is therefore estimated to be 0.58 ± 0.04.

This density result is slightly smaller than both the value of 1.71 ± 0.08 derived by Sicardy et al. (2006) and the value of 1.72 ± 0.15 derived by Gulbis et al. (2006). The radius measurements in each of these works are similar enough that the density difference results primarily from our using Buie et al.'s (2006) value of 1.52 ± 0.064 x $10^{21}$ kg for Charon's mass. Notably, the C313.2 occultation has constrained Charon's radius to the extent that the remaining error bars on its density are overwhelmingly the results of errors in the mass. Table 3 provides a comparison of the various density measurements and the masses used to calculate them.

## 5. DISCUSSION

Excluding both the Gemini South points and the SOAR immersion point, the remaining 13 occultation time measurements all closely correspond to the adopted figures at levels consistent with their formal errors. This agreement allows us to use the root mean square (rms) residual of 1.12 km from Fit 2 as an approximate upper limit on the rms roughness of Charon's limb. However, the near-identical durations of the Gemini and SOAR chords indicate that there can be significant deviations from this rms roughness value.

If the discrepancies between the Gemini and SOAR chords and the fitted models based on the other stations are interpreted not to be the result of timing anomalies, this could indicate a surface irregularity at that location. Given the residual pattern, and that the Gemini South chord used NTP rather than GPS timing, it is reasonable to accept the SOAR chord as the one most likely to be properly registered in time. However, even under the assumption that the Gemini chord is incorrectly timed, and the SOAR chord is correct, an elliptical fit to the data excluding Gemini still yields an unacceptably high reduced chi square (Fit 6). This indicates that the durations of SOAR and Gemini chords jointly point to a deviation from the simple circular or elliptical fits in this area of the limb. From their residuals with respect to both the circular and elliptical fits, we can determine the size of a possible feature measured at the Gemini and SOAR immersion location.

As previously stated, the SOAR chord resulted in residuals of –7.1 km and 0.1 km for the best-fitting circular model (Fit 2). The close correspondence of the emersion point (0.1 km from the circular model) could indicate that the SOAR timing was indeed valid within its formal errors. The Gemini chord is consistent with this when shifted to align its emersion point with the SOAR emersion point (see Figure 2). If true, this would indicate a surface depression at the immersion point of ~ 7 km in depth. This size is near the high end of likely surface features, but within reason for perhaps a large impact crater. Stern (1992) reports that features larger than 10 km would relax over geologic time scales on Pluto, resulting from the surface strength. Charon, being less massive, could therefore preserve a feature as large as 7 km for significant time if it has similar surface strength properties.

The calculated rock mass fraction of 0.58 ± 0.04, though lower than that reported by Gulbis et al. (2006) and Sicardy et al. (2006), is still larger than the maximum predicted by Charon formation models involving simple solar nebula condensation (McKinnon et al. 1997). This rock mass fraction points to a collisional formation scenario for the Pluto-Charon system, which would provide a means of losing ice mass through violent volatile escape. The result that Charon's bulk density is less than that of Pluto (Table 4) is consistent with this scenario when one or both of the parent collision objects were differentiated (McKinnon et al. 1997). The most likely scenario in this case is that of a low-velocity, oblique, two-body (both differentiated) collision resulting in Charon coalescing from a debris disk rather than being a surviving impactor (Canup 2005).

INSERT TABLE 4

The results from the C313.2 occultation reinforce the effectiveness of this method in exploring small bodies in the outer solar system. For comparison, the radii, bulk density and atmospheres for Triton, Pluto, Charon, and the three largest currently known Kuiper Belt objects are listed in Table 4. This comparison suggests that the larger Kuiper Belt objects could indeed have significant atmospheres (Elliot & Kern 2003) at appropriate portions of their orbits, which could be detected with stellar occultation observations. $2003UB_{313}$ for instance, is larger than Pluto, Charon, and possibly even Triton. If its formation allowed it to retain sufficient volatiles [methane has been detected on both $2003UB_{313}$ and $2005FY_9$ (Brown et al. 2005b; Licandro et al. 2005) and water has been detected on $2003EL_{61}$ (Trujillo et al. 2006)], it could likely support an atmosphere when it approaches the solar distance of Pluto. Unfortunately, $2003UB_{313}$ was discovered near the aphelion of its orbit at 97 AU from the sun and will not approach perihelion for some centuries. However, its mere existence provides hope for future discovery of objects nearer to perihelion that could support atmospheres that could be studied now.

## 6. CONCLUSIONS

We have measured a mean radius for Charon of 606.0 ± 1.5 km, which implies a bulk density of 1.63 ± 0.07 g/cm$^3$. Our analysis provides an upper limit on overall surface roughness for Charon ~1.1 km, with an overall planetary oblateness of 0.006 ± 0.003 (86% confidence level). However, there are strong indications that significant local deviations from these overall values exist.

Further refinement of these results can be obtained by future occultation observations at different Charon aspects. Many simultaneous chords would be needed to improve on these results and quantify local features on Charon’s surface—barring the existence of further striking, large-scale features that could be readily observed such as that indicated by the discrepancy between the Gemini South and SOAR chords and the rest of the data set.

Additionally, the results from this work could be used to fix Charon's radius in re-analyses of the Pluto-Charon mutual events, which would further constrain the derived value for the radius of Pluto’s visible disk at the epoch of the mutual events.

With the 2006 January 19 launch of the New Horizons mission, we can expect more definitive results about the character of Charon's topography during its flyby in 2015.

We would like to thank Gerard van Belle for his stellar diameter measurements of C313.2. This work was supported in part by NASA grants NNG04GE48G, NNG04GF25G, and NNH04ZSS001N.

REFERENCES

Anderson, J. D., Asmar, S. W., Campbell, J. K., Jacobson, R. A., Kushner, T. P., Kurinski, E. R., Lau, E. L., & Morabito, D. D. 1992. in Neptune and Triton Conference Proceedings, Gravitational parameters for Neptune and Triton (Tucson, Arizona), 1

Baier, G., & Weigelt, G. 1987, A&A, 174, 295

Bevington, P. R., & Robinson, D. K. 1992, Data Reduction and Error Analysis for the Physical Sciences (2nd ed.; New York: McGraw-Hill Inc.)

Bonneau, D., & Foy, R. 1980, A&A, 92, L1

Brown, M. E., Trujillo, C. A., & Rabinowitz, D. L. 2005a, International Astronomical Union Circular 8577

---. 2005b, ApJL, 635, L97

Buie, M., Grundy, W., Young, E. F., Young, L. A., & Stern, S. A. 2006, preprint (astro-ph/0512491)

Canup, R. M. 2005, Science, 307, 546

Chamberlin, A. B. 2005, JPL Horizons: Solar System Dynamics, JPL / NASA, http://ssd.jpl.nasa.gov/

Deeths, D., & Brunette, G. 2001, in (Palo Alto: Enterprise Engineering, Sun Microsystems, Inc.), 7

Elliot, J. L., & Kern, S. D. 2003, Earth, Moon, and Planets, 92, 375

Elliot, J. L., & Olkin, C. B. 1996, in Annual Review of Earth and Planetary Sciences, eds. G. W. Wetherill, A. L. Albee, & K. C. Burke (Palo Alto: Annual Reviews Inc.), 89

Elliot, J. L., et al. 2000a, Icarus, 148, 347

Elliot, J. L., Person, M. J., & Qu, S. 2003, AJ, 126, 1041

Elliot, J. L., Strobel, D. F., Zhu, X., Stansberry, J. A., Wasserman, L. H., & Franz, O. G. 2000b, Icarus, 143, 425

Elliot, J. L., & Young, L. A. 1991, Icarus, 89, 244

---. 1992, AJ, 103, 991

Gangestad, J. W. 2006, B.S. in Department of Astronomy, The study of Pluto, Charon, and Kuiper Belt objects through stellar occultations, (Williams College), in preparation

Gulbis, A. A. S., et al. 2006, Nature, 439, 48

Licandro, J., Pinilla-Alonso, N., Pedani, M., Oliva, E., Tozzi, G. P., & Grundy, W. 2005, A&A, 445, L35

McDonald, S. W., & Elliot, J. L. 2000, AJ, 120, 1599

McKinnon, W. B., Simonelli, D. P., & Schubert, G. 1997, in Pluto and Charon, eds. S. A. Stern, & D. J. Tholen (Tucson: University of Arizona Press), 295

Null, G. W., Owen, W. M., & Synnott, S. P. 1993, AJ, 105, 2319

Olkin, C. B., Wasserman, L. H., & Franz, O. G. 2003, Icarus, 164, 254

Person, M. J. 2006, Ph. D. in Earth, Atmospheric, and Planetary Sciences, Stellar Occultation Studies of Charon, Pluto, and Triton, (Massachusetts Institute of Technology), in preparation

Rabinowitz, D. L., Barkume, K. M., Brown, M. E., Roe, H., Schwartz, M., Tourtellotte, S., & Trujillo, C. A. 2006, ApJ, submitted (astro-ph/0509401)

Reinsch, K., Burwitz, V., & Festou, M. C. 1994, Icarus, 111, 503–512

Sicardy, B., et al. 2006, Nature, 439, 52

Souza, S. P., Babcock, B. A., Pasachoff, J. M., Gulbis, A. A. S., Elliot, J. L., & Person, M. J. 2006, (in preparation)

Stern, S. A. 1992, Annual Reviews of Astronomy and Astrophysics, 30, 185

Tholen, D. J., & Buie, M. W. 1990, BAAS, 22, 1129

---. 1997, in Pluto and Charon, eds. S. A. Stern, & D. J. Tholen (Tucson: Univ. of Arizona Press), 193

Thomas, P. C. 2000, Icarus, 148, 587

Trujillo, C. A., Brown, M. E., Barkume, K. M., Schaller, E. L., & Rabinowitz, D. L. 2006, ApJ, submitted (astro-ph/0601618)

Tytell, D. 2005, S&T, 110, 28

USNO. 2003, The Astronomical Almanac for the Year 2005 (Washington, D.C.: U.S. Government Printing Office)

Walker, A. R. 1980, MNRAS, 192, 47p

Weaver, H. A., & Stern, S. A. 2005, IAUC, No. 8625

Young, E. F., & Binzel, R. P. 1994, Icarus, 108, 219

Young, L. A., Olkin, C. B., Young, E. F., & French, R. G. 2005, International Astronomical Union Circular 8570

Zacharias, N., Urban, S. E., Zacharias, M. I., Wycoff, G. L., Hall, D. M., Monet, D. G., & Rafferty, T. J. 2004, AJ, 127, 3043

TABLE 1
SUMMARY OF 2005 JULY 11 PUBLISHED CHARON OCCULTATION CHORDS

| Chord | Telescope Diameter (m) | Location (E Longitude, S Latitude) | Occultation Times (UT) | |
|---|---|---|---|---|
| | | | Immersion | Emersion |
| San Pedro de Atacama[a] | 0.5 | –68º 10º 48.2’’, 22º 57’ 08.4’’ | 3:36:20.98 ± 0.18 | 3:36:28.30 ± 0.30 |
| Cerro Armazones[b] | 0.84 | –70º 11’ 46’’, 24º 35 ’52’’ | 3:36:16.99 ± 0.04 | 3:36:54.28 ± 0.03 |
| Paranal[a] | 8.2 | –70º 24’ 07.9’’, 24º 37’ 31.0’’ | 3:36:18.09 ± 0.04 | 3:36:55.40 ± 0.05 |
| Las Campanas – du Pont[b] | 2.5 | –70º 42’ 13’’, 29º 00’ 26’’ | 3:36:13.792 ± 0.005 | 3:37:10.609 ± 0.004 |
| Las Campanas – Clay[b] | 6.5 | –70º 42’ 33’’, 29º 00’ 51’’ | 3:36:13.774 ± 0.001 | 3:37:10.563 ± 0.002 |
| Cerro Pachón – Gemini South[b] | 8.0 | –70º 43’ 24’’, 30º 13’ 42’’ | 3:36:15.50 ± 0.15 | 3:37:10.55 ± 0.15 |
| Cerro Pachón – SOAR[c] | 4.2 | –70º 44’ 1.4’’, 30º 14’ 16.8’’ | 3:36:16.19 ± 0.01 | 3:37:11.26 ± 0.01 |
| El Leoncito[a] | 2.15 | –69º 17’ 44.9’’, 31º 47’ 55.6’’ | 3:36:15.03 ± 0.16 | 3:37:02.98 ± 0.08 |

[a] Sicardy et al. (2006).

[b] Gulbis et al. (2006).

[c] Young et al. (2005).

TABLE 2
CHARON FIGURE FIT RESULTS

| Fit | Data Selection | Mean Radius (km)[a] | Observed Oblateness | Position Angle (deg) | $f_0$ (km)[b] | $g_0$ (km)[b] | Reduced Chi Square |
|---|---|---|---|---|---|---|---|
| 1-Circular | All Data | 606.03 ± 0.20 | - | - | 685.81 ± 0.44 | –783.74 ± 3.75 | 72.4 |
| **2-Circular**[c] | **Gemini and SOAR omitted** | **606.01 ± 0.02** | **-** | **-** | **686.18 ± 0.06** | **–787.62 ± 0.55** | **1.27** |
| 3-Circular | Gemini omitted | 606.06 ± 0.18 | - | - | 685.66 ± 0.40 | –782.33 ± 3.38 | 67.7 |
| 4-Elliptical | All data | 599.13 ± 5.48 | 0.027 ± 0.010 | 95.5 ± 7.8 | 684.87 ± 0.47 | –776.62 ± 3.65 | 33.5 |
| **5-Elliptical**[c] | **Gemini and SOAR omitted** | **604.52 ± 1.83** | **0.006 ± 0.003** | **71.4 ± 10.4** | **685.98 ± 0.17** | **–785.54 ± 1.60** | **0.89** |
| 6-Elliptical | Gemini omitted | 598.84 ± 4.52 | 0.029 ± 0.008 | 97.9 ± 5.7 | 684.79 ± 0.36 | –776.47 ± 2.77 | 31.5 |

[a] For elliptical fits, this is the geometric mean of the semi-major and -minor axes.

[b] These positions are a measure of the offset between the observed center of the *(f,g)* coordinate system and the expected location of Charon at the time of the occultation. Thus, they provide a measure of the astrometric uncertainties in the difference between the occulted star's position and Charon's ephemeris. The $f_0$ and $g_0$ values given in Fit 1 result in astrometric offsets of 0.03145 ± 0.00002 arcsec in RA and –0.03594 ± 0.00017 arcsec in Dec (Charon's offset – star's offset). However, it is not possible to disentangle the individual contributions to these errors from the UCAC catalog and Charon's ephemeris without making further assumptions.

[c] Rows in boldface represent our adopted, best-fit circular and elliptical solutions.

TABLE 3
COMPARISON OF CHARON RADII AND DENSITY MEASUREMENTS

| Source | Method | Mean Charon Radius (km) | Charon Density[a] ($g/cm^3$) |
|---|---|---|---|
| **This Work** | occultation[b] | **606.0 ± 1.5** | **1.63 ± 0.07**[c] |
| Gulbis et al. (2006) | occultation[b] | 606 ± 8 | 1.72 ± 0.15[d] |
| Sicardy et al. (2006) | occultation[e] | 603.6 ± 1.4 | 1.71 ± 0.08[f] |
| Young et al. (2005) | occultation[g] | > 589.5 ± 2 | – |
| Reinsch et al. (1994) and Young et al. (1994) | mutual events[h] | 591 ± 5 – 628 ± 21 | – |
| Elliot and Young (1991) | occultation[g] | > 601.5 | – |
| Baier and Weigelt (1987) | speckle interferometry | 525–760 | – |
| Walker (1980) | occultation[g] | > 600 | – |
| Bonneau and Foy (1980) | speckle interferometry | 1000 ± 100 | – |

[a] As the first reliable Charon mass measurement was published by Null et al. (1993) we do not list earlier density estimates.

[b] Both circular and elliptical solutions were considered in calculating the error bar.

[c] Assuming Charon's mass = 1.520 ± 0.064 x $10^{21}$kg (Buie et al. 2006).

[d] Assuming Charon's mass = 1.60 ± 0.12 x $10^{21}$kg (Olkin et al. 2003).

[e] Circular solutions were considered in calculating the error bar.

[f] Assuming Charon's mass = 1.58 ± 0.07 x $10^{21}$kg [R.A. Jacobson, personal communication to Sicardy et al. (2006)].

[g] Lower limit determined from a single occultation chord.

[h] Lower and upper limits from the various published mutual event analyses using various assumptions about limb darkening. See Tholen and Buie (1997) for further details.

TABLE 4
PROPERTIES OF SMALL BODIES IN THE OUTER SOLAR SYSTEM

| Object | Radius (km) | Mass ($10^{24}$ g) | Bulk Density (g/cm$^3$) | Atmosphere |
|---|---|---|---|---|
| Triton | 1353.4 ± 0.9[a] | 21.398 ± 0.053[b] | 2.0607 ± 0.0066 | ~ 19 μbar $N_2$[c] |
| $2003UB_{313}$ | 1405 – 5625[d] | ? | ? | ? |
| Pluto | 1175 ± 25[e] | 13.050 ± 0.065[f] | 1.92 ± 0.12 | 5–11 μbar $N_2$[g] |
| $2005FY_9$ | 775 – 1550[h] | ? | ? | ? |
| $2003EL_{61}$ | 855 – 1245[i] | 4.21 ± 0.10[i] | 2.60 – 3.34[i] | ? |
| Charon | 606.0 ± 1.5[j] | 1.520 ± 0.064[f] | 1.63 ± 0.07 | <0.11 μbar $N_2$[k] |

[a] Thomas (2000).

[b] Anderson et al. (1992).

[c] Reported surface pressure was measured to increase between the 1995 and 1997 occultations (Elliot et al. 2000a; Elliot et al. 2000b).

[d] Radius range is based upon absolute magnitude ($H_R$ = –1.48) from Brown et al. (2005) assuming an *R* albedo range of 0.04 – 0.64.

[e] This radius range is a consensus value from many different observations. See Tholen and Buie (1997) for further details.

[f] Buie et al. (2006).

[g] The range in surface pressures is due to a choice of atmospheric models (Elliot et al. 2003).

[h] Radius range is based upon absolute magnitude ($H_V$ = –0.1) from Licandro et al. (2005) assuming a *V* albedo of 0.2 – 0.8.

[i] Rabinowitz *et al*. (2006). This object is thought to be an elongated ellipsoid so the listed radius range represents half the total length.

[j] This work.

[k] Upper limit for a nitrogen atmosphere, based on non-detection during C313.2 event. The larger of the two derived upper limits is presented (Gulbis et al. 2006; Sicardy et al. 2006).

FIGURE CAPTIONS

**FIG. 1. – Charon Figure Solution**: Occultation immersion and emersion points are plotted in the (*f,g*) plane for all stations listed in Table 1. Designations (a,b,c) refer to data from Sicardy et al. (2006), Gulbis et al. (2006), and Young et al. (2006).. The best-fitting circular solution (Fit 2 from Table 2) is plotted as a solid circle, while the best-fitting elliptical solution (Fit 5 from Table 2) is dashed. Points from co-located stations (e.g. du Pont and Clay) appear on top of each other at this scale. The reported formal error bars are smaller than the plotted points. (See Figure 2 for Gemini South error bars.) Note the clear deviations of the SOAR and Gemini points from the best-fit solutions.

**FIG. 2. – Expanded View of Gemini South and SOAR Residuals**: An expanded view of Charon's limb solution from Figure 1 is plotted along with points from four of the stations. The best-fitting circular solution (Fit 2 from Table 2) is plotted with solid lines, while the best-fitting elliptical solution (Fit 5 from Table 2) is dashed. Both the Las Campanas stations (Clay and du Pont) and the Cerro Pachón stations (Gemini South and SOAR) are plotted according to their reported occultation times. Error bars are shown for Gemini South, but all other stations reported error bars are smaller than the size of the points shown. Note that on the emersion limb (right side), the SOAR point falls almost directly on the model curves, while on the immersion limb it is 7 km away. If the Gemini chord were shifted (by a systematic timing error, for example), its points would lie almost directly on the SOAR points at this scale. The shaded area on the left panel is suggestive of the size of a local figure anomaly (such as a large impact crater) that would be needed to account for the residuals for these two chords.

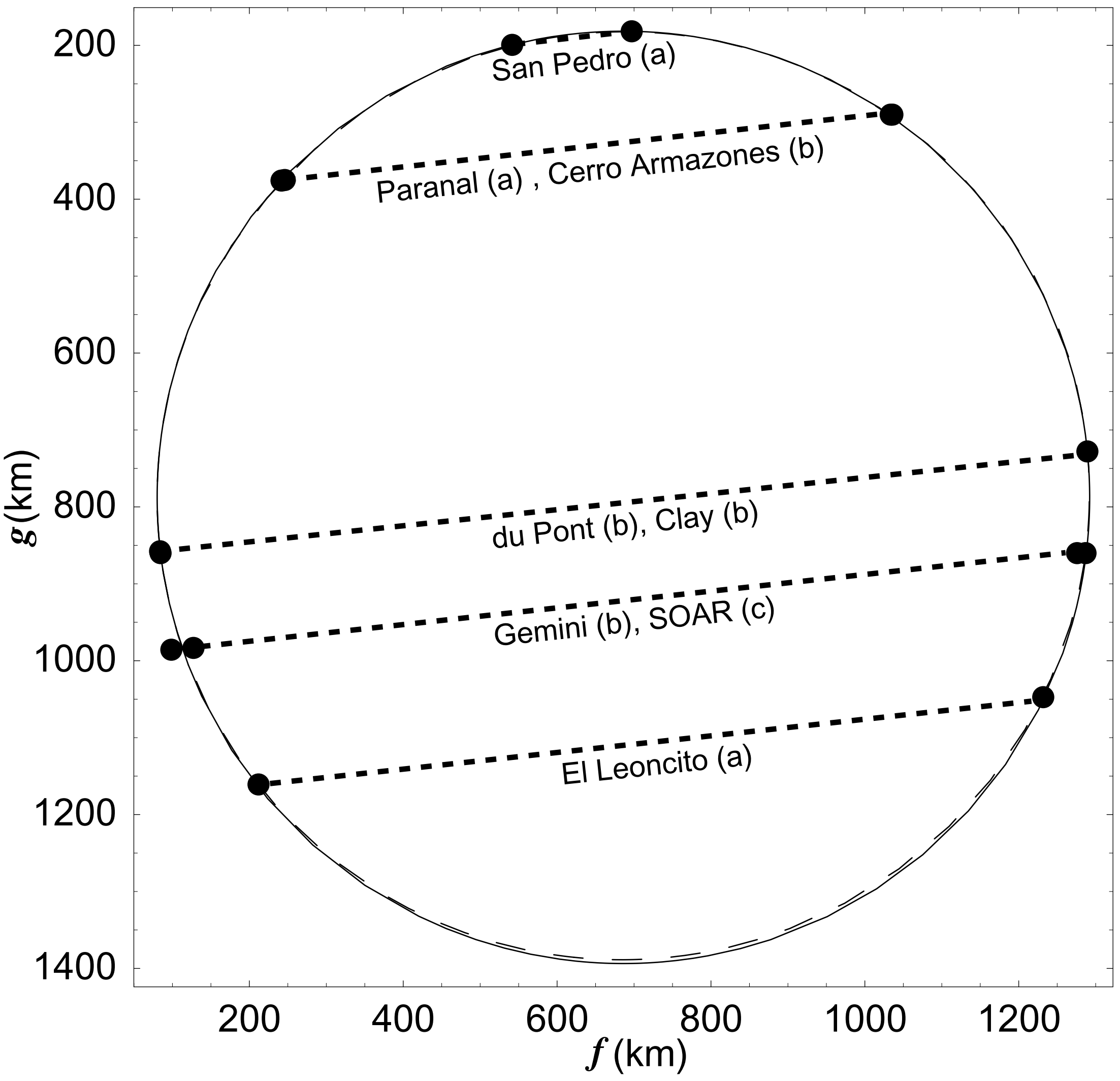

**Figure 1: Person,** ***et al.***

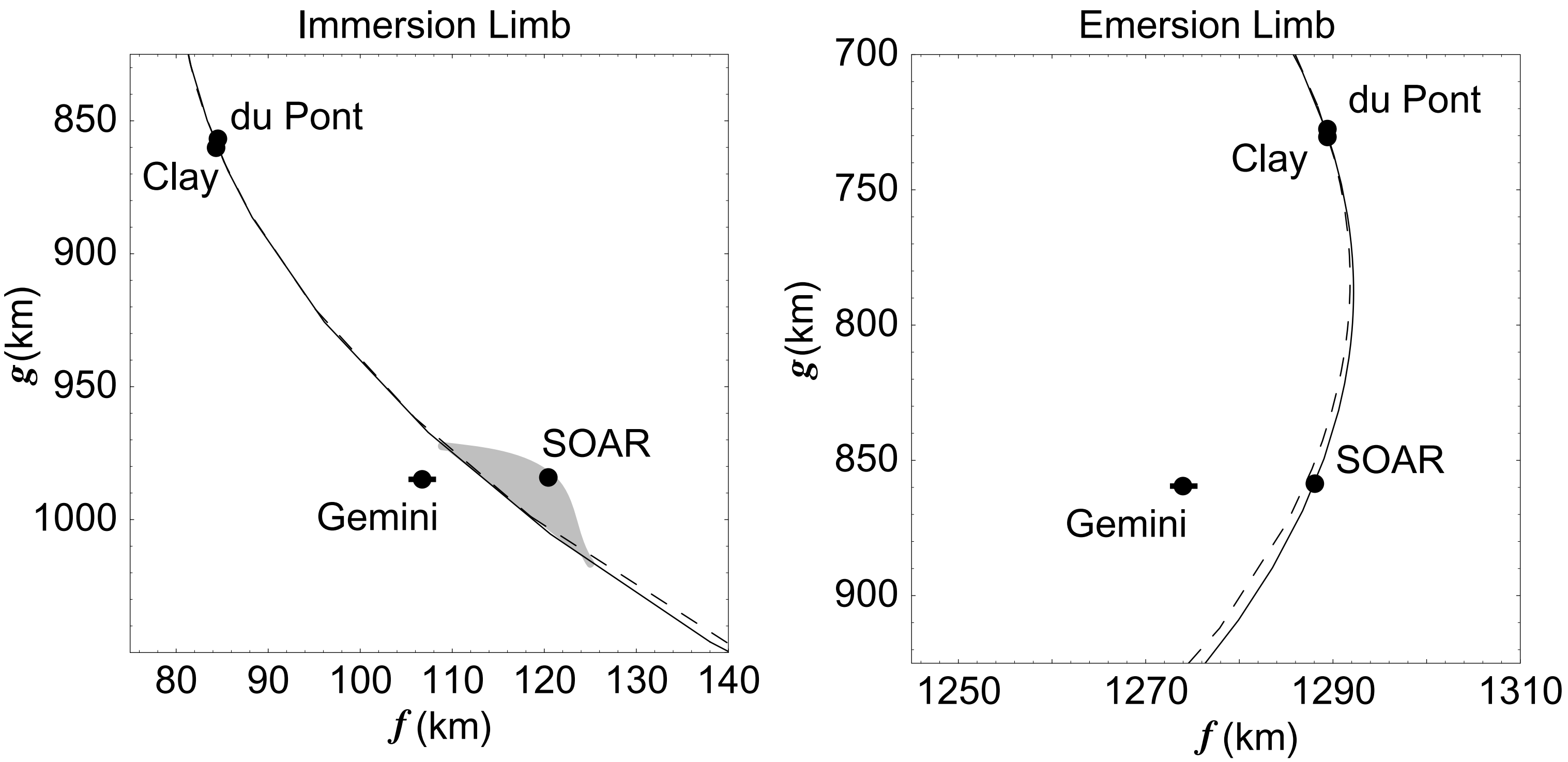

**Figure 2: Person,** ***et al.***